\documentclass[10pt,a4paper]{article}

\pdfoutput=1

\usepackage{epsfig}
\usepackage[usenames]{color}
\usepackage{url} 
\usepackage{hyperref}

\newcommand{\be}{\begin{equation}}
\newcommand{\ee}{\end{equation}}
\newcommand{\bea}{\begin{eqnarray}}
\newcommand{\eea}{\end{eqnarray}}

\newcommand{\gapp}{\mathrel{\raise.3ex\hbox{$>$}\mkern-14mu  
\lower0.6ex\hbox{$\sim$}}}
\newcommand{\lapp}{\mathrel{\raise.3ex\hbox{$<$}\mkern-14mu  
\lower0.6ex\hbox{$\sim$}}}
\newcommand{\maxa}{\mathrel{\raise.1ex\hbox{${\rm max}$}\mkern-14mu  
\lower0.7ex\hbox{$a$}}}
\newcommand{\mina}{\mathrel{\raise.1ex\hbox{${\rm min}$}\mkern-14mu  
\lower0.7ex\hbox{$a$}}}

\newcommand{\ECM}{E_{\rm CM}}
\begin{document}

\begin{center}

{\Huge Manual of BlackMax}

\vspace*{2mm}

{\large A black-hole event generator with rotation, recoil, split branes, and brane tension}

\vspace*{5mm}

{\large Version 2.00} 

\vspace*{5mm}

De-Chang Dai$^1$, Cigdem Issever$^2$, Eram Rizvi$^3$, Glenn Starkman$^4$, Dejan Stojkovic$^1$, Jeff Tseng$^2$

\vspace*{3mm}

$^1${\it Department of Physics, SUNY at Buffalo, Buffalo NY 14260-1500,USA}

$^2${\it University of Oxford, Oxford,  UK}

$^3${\it Queen Mary, University of London, London, UK}

$^4${\it CERCA and ISO, Department of Physics, Case Western Reserve University, Cleveland OH 44106-7079, USA}

\end{center}

\vspace*{5mm}


This is the users manual of the black-hole event generator BlackMax~\cite{Dai:2007ki},
which simulates the experimental signatures of microscopic and 
Planckian black-hole production and evolution at proton-proton, 
proton-antiproton and electron-positron colliders in the context of 
brane world models with low-scale quantum gravity.
The generator is based on phenomenologically realistic models free of 
serious problems that plague low-scale gravity. 
It includes all of the black-hole gray-body factors known to date 
and incorporates the effects of black-hole rotation, splitting between 
the fermions, non-zero brane tension and black-hole recoil due to Hawking radiation
(although not all simultaneously).
The main code can be downloaded from \cite{code}. 


\section{Introduction and New Features}
\label{sec:features}

BlackMax is a very versatile semi classical and quantum black-hole
generator which simulates a number of different extra dimension models and black-hole
evolution scenarios. It also gives the user the possibility to set many
of the parameters which influence the formation and decay of black
holes. The manual
intends to explain the various parameters the user can set.\\


After the release of BlackMax version 1.01 we received comments, bug-reports
and requests for new features from the user's community, which we
tried to incorparate as much as possible. The new features of
BlackMax version 2.00 are listed below and will be discussed in the
upcoming sections in more detail.

\begin{enumerate}

\item BlackMax is able to simulate black-hole production in proton-proton,
protron-anti-proton and electron-positron collisions. 

\item The user is able to choose among different Planck scale
conventions or use his/her own convention.

\item Graviton emission can be simulated. 

\item The user can define if baryon, lepton numbers or flavour should
be conserved or not. {\bf In case the user does not conserve baryon numbers
BlackMax events cannot be hadronised with Pythia or HERWIG.} 

\item The parton density functions of the
LHAPDF package~\cite{weblhapdf} are now included into BlackMax.

\item The user can choose to use the Yoshino-Rychkov cross section enhancement factors
and simulate the energy loss before the formation of the event horizon as
described in ~\cite{Yoshino:2005hi}. These corrections to the cross section 
and the energy loss are only applicable for non-tension models and BlackMax 
will not apply any corrections or modifications to the energy loss if the user 
turns the Yoshino-Rychkov corrections on for a positive tension brane scenario. 

\item The default BlackMax output is in the LHA format~\cite{Boos:2001cv}. 

\item Graviton and photon emission is now also included into the
balding phase before the black-hole is formed. If the user has turned
on graviton emission and choses to
use the Yoshino-Rychkov suppression factors or sets the mass loss factor to a non-zero value, BlackMax sheds
this energy/mass by emitting two gravitons before the
formation of the black-hole. If the user has disabled graviton
emission BlackMax will shed this energy by emitting two photons.   

\item If the user simulates the split-fermion scenario
BlackMax takes into account the position of the black-hole remnant
when it calculates which particles it should decay into during
the final burst phase.

\item Users should note that the output format has changed with
respect to the description in~\cite{Dai:2007ki} and is described in
section~\ref{subsec:output}.

\end{enumerate}


\section{Installation}
\label{sec:installation}

The most up-to-date source code and TarBall can be downloaded from:\\
\\
   \url{http://projects.hepforge.org/blackmax/}\\
\\
Having downloaded the zipped tar file you will need to unzip it, then
extract the files and make the executable:\\
\\
{\tt 
gunzip 	 BlackMax-2.00.tar.gz\\
tar -xvf BlackMax-2.00.tar\\
}
\\
Before compilation you will need to check the compiler version of gcc you are 
using. This is because the latest gcc compiler version (4.1.2) has changed the 
names of some system libraries needed to compile Fortran with C code. The 
download is configured to use gcc version 4. If you have an older gcc version 
e.g. 3.4.6 then you will need to modify the BlackMax Makefile.
You can check your compiler version by doing the following:\\
\\
{\tt gcc --version}
\\
Which will generate output like this:\\
\\
{\tt gcc (GCC) 3.4.6 20060404 (Red Hat 3.4.6-10)\\
Copyright (C) 2006 Free Software Foundation, Inc.\\
...}\\
\\
You will need to change the Fortran system library names in the 
Makefile in case your compiler version is an older one. Do this by
uncommenting the following lines in the Makefile\\   
\\
{\tt   
   F77LIB =g2c\\
   F77COMP=g77}\\
\\
You are now ready to compile BlackMax.

There are three different ways to run BlackMax:
\begin{enumerate}
  \item in standalone mode - no additional libraries required
  \item accessing PDFs from LHAPDF
  \item accessing PDFs from LHAPDF and simultaneous hadronisation from Pythia
\end{enumerate}

To produce the exectuable for each method simply requires a different
compilation/linking step and is described below. In all three
options the default format of the event output is the Les Houches
Accord format~\cite{Boos:2001cv}. This text file can be used as input
to HERWIG/Pythia to hadronise the BlackMax events at a later
date\footnote{Only if baryon numbers are conserved, please see
section~\ref{subsec:input}, bullet item 29.}

\subsection{To Run in Standalone Mode}

In this version the proton parton densities are taken from CTEQ6m which
are packaged with BlackMax. After unpacking simply do:\\
\\
{\tt gmake BlackMaxOnly}\\
\\
modify parameter.txt to pick one of the 41 CTEQ6m PDF sets that has been bundled 
with BlackMax, for example:\\
\\
        {\tt choose$\_$a$\_$pdf$\_$file(200$\_$to$\_$240$\_$cteq6)Or$\_>$10000$\_$for$\_$LHAPDF}\\
        200\\
\\
Then run the executable\\
\\
{\tt BlackMax $>\&!$ out}
\\

\subsection{To Run with LHAPDF}
\label{subsec:lhapdf}
This version of the executable uses the proton parton densities from the LHAPDF library which you will need to
download and install from here:\\  
\\
\url{http://projects.hepforge.org/lhapdf/}\\
\\
For LHAPDF please ensure you install the package in a directory where 
you have write permission. You can do this by specifying an 
installation directory - for more info see the LHAPDF manual.
Edit the BlackMax Makefile and insert the library locations. Ensure
that your ${\tt LD\_LIBRARY\_PATH}$ environment variable includes the location
of your newly built LHAPDF library:\\
\\
export ${\tt LD\_LIBRARY\_PATH=\$LD\_LIBRARY\_PATH:/data/rizvi/atlas/lhapdf-5.3.0/lhapdf/lib}$\\
export ${\tt LHAPATH=/data/rizvi/atlas/lhapdf-5.3.0/lhapdf/share/lhapdf/PDFsets}$\\
\\
Ensure you have chosen a valid PDF set in parameter.txt, for example to choose the LHAPDF
partons from the H1 PDF2000 fit of HERA data:\\
\\
        {\tt choose$\_$a$\_$pdf$\_$file(200$\_$to$\_$240$\_$cteq6)Or$\_>$10000$\_$for$\_$LHAPDF}\\
        70050\\

After unpacking the source files do:\\
\\
{\tt gmake BlackMax}\\
\\
Once the executable is created you can run the program:\\
\\
{\tt BlackMax $>\&!$ out}

\subsection{To Run with Simultaneous Pythia Hadronisation}

In order the hadronise the events during the generation job BlackMax comes with an interface to
Pythia. The user can only run BlackMax simultaneously with Pythia if
s/he is conserving baryon numbers (see section~\ref{subsec:input}, bullet item
29). In order to generate fully hadronised events you will need to 
download and install the latest versions of LHAPDF and PYTHIA. They
are available at:\\
\\
\url{http://www.hepforge.org/downloads/pythia6}\\
\url{http://www.hepforge.org/downloads/lhapdf}\\
\\
BlackMax has been tested wth Pythia 6.4.10 and LHAPDF 5.3.0.
Install Pythia and LHAPDF according to the instructions. With Pythia,
create the libraries, but remove the following dummy routines 
from Pythia:\\
\\
	upinit.f\\
	upevnt.f\\
	pdfset.f\\
	structm.f\\
\\

You will also need to remove the mention of the pdfset.f routine from the Pythia Makefile. 
The four routines above are all dummy routines which actually exist in LHAPDF.
Edit the BlackMax Makefile and insert the library locations. Ensure
that your ${\tt LD\_LIBRARY\_PATH}$ environment variable includes the location
of your newly built Pythia and LHAPDF libraries:\\
\\
export ${\tt LD\_LIBRARY\_PATH=\$LD\_LIBRARY\_PATH:/data/rizvi/atlas/lhapdf-5.3.0/lhapdf/lib}$\\
export ${\tt LHAPATH=/data/rizvi/atlas/lhapdf-5.3.0/lhapdf/share/lhapdf/PDFsets}$\\
\\
Then create the BlackMax executable using the target "all" which will
link to the Pythia and LHAPDF libraries:\\
\\
{\tt gmake all}\\
\\
Ensure you have chosen a valid PDF set in parameter.txt, for example:\\
\\
        ${\tt choose\_a\_pdf\_file(200\_to\_240\_cteq6)Or\_>10000\_for\_LHAPDF}$\\
        $10050$\\
\\
then run the exectuable:\\
\\
{\tt BlackMax $>\&!$ out}
\\


\section{Black-Hole Production} 
\label{sec:bhproduction} 
\indent 
We assume that the fundamental quantum-gravity energy scale $M_*$  
is not too far above the electroweak scale.  
Consider two particles colliding with a center-of-mass energy $E_{CM}$.  
They will also have an angular momentum $J$ in their center-of-mass (CM) frame. 
By the hoop conjecture~\cite{Thorne:1994xa}, if the impact parameter, $b$,  
between the two colliding particles is smaller than  
the diameter of the horizon of a $(d+1)$-dimensional black-hole  
(where $d$ is the total number of space-like dimensions) of mass $M=E_{CM}$ and   
angular momentum $J$, 
\be 
b < 2 r_h(d,M,J) , 
\ee 
then a black-hole with $r_{h}$ will form. 
The cross section for this process is approximately equal to the 
interaction area $\pi (2r_h)^2$. \\
 
In Boyer-Lindquist coordinates, 
the metric for a $(d+1)$-dimensional rotating black-hole 
(with angular momentum parallel to the $\hat{\omega}$ in the rest frame of the   
black-hole) is: 
\begin{eqnarray} 
\label{eqn:bhmetric} 
ds^2 &=&\left( 1-\frac{\mu r^{4-d}}{\Sigma (r,\theta)}\right)dt^{2} 
\nonumber \\ 
&-& \sin^{2}\theta \left(r^{2}+a^{2}\left( 
+ \sin^{2}\theta \frac{\mu r^{4-d}}{\Sigma (r,\theta)}\right)\right) 
d\phi ^{2} \nonumber \\ 
&+& 2 a \sin^{2}\theta \frac{\mu r^{4-d}}{\Sigma (r,\theta )} dt d\phi 
- \frac{\Sigma(r,\theta)}{\Delta} dr^{2} \nonumber \\ 
&-& \Sigma(r,\theta) d\theta ^{2} - r^{2}cos^2\theta 
d^{d-3}\Omega 
\end{eqnarray} 
where $\mu$ is a parameter related to mass of the black-hole, while  
\begin{eqnarray} 
&&\Sigma =r^2+a^{2}cos^{2}\theta \\ 
{\rm and}&& \nonumber \\ 
&& \Delta=r^2+a^{2}-\mu r^{4-d} . 
\end{eqnarray} 
The mass of the black-hole is 
\begin{equation} 
M=\frac{(d-1) A_{d-1}}{16\pi G_d}\mu , 
\end{equation} 
and 
\begin{equation} 
J=\frac{2Ma}{d-1} 
\end{equation} 
is its angular momentum. Here, 
\be 
A_{d-1} = \frac{2 \pi^{d/2}}{\Gamma(d/2)} 
\ee 
is the hyper-surface area of a $(d-1)$-dimensional unit sphere. 
The higher-dimensional gravitational constant $G_d$ is defined as 
\begin {equation} 
G_{d}=\frac{(2\pi)^{d-4}}{4 M_{\star}^{d-1}} . \label{eqn:G}
\end{equation} 
 
(The user is able to choose other Planck scale conventions than the
one in equation~\ref{eqn:G}, see Section~\ref{sec:input_output}). 
The horizon occurs when $\Delta =0$. That is at a radius given 
implicitly by 
\begin{equation} 
\label{eqn:horizon} 
r^{(d)}_h=\left[\frac{\mu}{1+(a/r^{(d)}_h)^{2}}\right]^{\frac{1}{d-2}} 
=\frac{r^{(d)}_s}{\left[1+(a/r^{(d)}_h)^{2}\right]^{\frac{1}{d-2}}}. 
\end{equation} 
Here 
\be 
\label{eqn:defrSchwarzshild} 
r^{(d)}_s\equiv\mu ^{1/(d-2)} 
\ee 
is the Schwarzschild radius of a $(d+1)$-dimensional black-hole, 
{\it i.e.} the horizon radius of a non-rotating black-hole. 
Equation \ref{eqn:defrSchwarzshild} can be rewritten as: 
\begin{equation} 
\label{eqn:horizon2} 
r^{(d)}_s(\ECM,d,M_*)=k(d)M_{*}^{-1}[\ECM/M_{*}]^{1/(d-2)}, 
\end{equation} 
where 
\begin{equation} 
\label{eqn:horizon3} 
k(d)\equiv\left[2^{d-3}{\pi}^{(d-6)/2}\frac{\Gamma[d/2]}{d-1}\right]^{1/(d-2)} . 
\end{equation} 
 
The Hawking temperature of a black-hole is  
\be 
\label{eqn:THawking} 
T_{H} = \frac{d-2}{4\pi r_h} .
\ee 
 
If two highly relativistic particles collide with center-of-mass energy 
$E_{CM}$, and impact parameter $b$, then their angular momentum 
in the center-of-mass frame before the collision is $L_{in}=b E_{CM}/2$.  
Suppose for now that the black-hole that is formed  
retains all this energy and angular momentum. 
Then the mass and angular momentum of the black-hole will be 
$M_{in}=\ECM$ and $J_{in}=L_{in}$. 
A black-hole will form if: 
\begin{equation} 
\label{bmax1} 
b < b_{\rm max} \equiv 2r^{(d)}_h (E_{CM}, b_{max}E_{CM}/2) \, . 
\end{equation} 
We see that $b_{max}$ is a function of both $E_{CM}$ and the number of 
extra dimensions. 
 
We can rewrite condition (\ref{bmax1}) as 
\begin{equation} 
\label{eqn:bmaxfinalform} 
b_{max}(E_{CM};d)= 
2\frac{r^{(d)}_s(E_{CM})}{\left[1+\left(\frac{d-1}{2}\right)^{2} 
\right]^{1 \over {d-2}}} \, ~\cite{Ida:2003gx}. 
\label{bmax2} 
\end{equation} 
 
There is one exception to this condition. In the case  
where we are including the effects of the brane tension, 
the metric (and hence gray-body factors) for a rotating 
black-hole are not known. In this case we 
consider only non-rotating black-holes. 
For the model with with non-zero tension brane, the radius of the black-hole is   
defined as  
\begin{equation} 
\label{eqn:r-tension} 
r_{h}=\frac{r_{s}}{B^{1/3}}, 
\end{equation} 
with $B$ the deficit-angle parameter which is inverse proportional to the   
tension of the brane $B=1-\frac{\lambda}{2\pi
M^{d-2}_{d}}$ (see equation 8 in~\cite{NKD}).
Therefore, for branes with tension  
\be 
b^{tension}_{max}(E_{CM},d) = 2 r^{(d)}_h(\ECM) . 
\ee 
Also, for branes with positive tension only the $d=5$ metric is known.

\section{Simulated Scenarios in BlackMax} \label{sec:graybodyfactors} 
\indent 

BlackMax is able to simulate several different extra dimension models,
black-hole scenarios.


\begin{itemize}
\item {\it Non-rotating black-hole on a tensionless brane:} 
For a non-rotating black-hole, we used previously known gray-body factors 
for spin $0,\frac{1}{2}$ and $1$ fields in the brane, and for spin $2$ fields 
({\it i.e.} gravitons) in the bulk.

\item {\it Rotating black-hole on a tensionless brane:} 
For rotating black-holes, we used known gray-body factors for spin 
$0,\frac{1}{2}$ and $1$ fields on the brane. The correct emission spectrum
for spin $2$ bulk fields is not yet known for rotating black-holes,
we currently do not allow for the emission of bulk gravitons from
rotating black-holes.

\item {\it Non-rotating black-holes on a tensionless brane with fermion brane 
splitting:}
In the split-fermion models, gauge fields can propagate through the bulk
as well as on the brane, so we have calculated gray-body factors for 
spin $0$ and $1$ fields propagating through the bulk,
but only for a non-rotating black-hole for the split-fermion model~\cite{Dai:2006dz}.

\item {\it Non-rotating black-holes on a non-zero tension brane:} 
The bulk gray-body factors for a brane with non-zero tension are affected by
non-zero tension because of the modified bulk geometry (deficit angle).
We have calculated gray-body factors for spin $0,1$ and $2$ fields 
propagating through the bulk, again only for the non-rotating black-hole for a 
brane with non-zero tension and $d=5$.

\item {\it Two particle final states:}
We use the same gray-body factors as a non-rotating black-hole to calculate 
the cross section of two-particle final states (excluding gravitons according to \cite{Meade}).
\end{itemize}


\section{The Black-Hole Formation}
The formation of the black-hole is a very non-linear and complicated process.  
BlackMax gives the user the possibility to set parameters which effect
the formation phase of the black-hole by hand or use corrections
calculated by Yoshino and Rychkov \cite{Yoshino:2005hi}.

If the user would like to set the formation parameters her/himself
BlackMax assumes that, before settling down to a stationary phase,  
a black-hole loses some fraction of its energy, linear and angular momentum.  
These losses are parameterize by three parameters:  
$1 - f_{E}$, $1-f_{P}$ and $1-f_{L}$.  
Thus the black-hole initial state that we actually evolve is characterized by  
\begin{eqnarray} 
\label{eqn:initlossfactors} 
E&=&E_{in}f_{E};\nonumber \\ 
P_z&=&P_{z\, in}f_{P};\\ 
J'&=&L_{in}f_{L}\nonumber ; 
\end{eqnarray} 
where $E_{in}$, $P_{z\, in}$ and $L_{in}$ are initial energy,  
momentum and angular momentum of colliding partons,  
while $f_{E}$, $f_{P}$ and $f_{L}$ are the fractions of  
the initial energy, momentum and angular momentum 
that are retained by the stationary black-hole.  



\section{Black-Hole Evolution in BlackMax} 
\label{sec:bhevolution} 
\indent 
 
The Hawking radiation spectra are calculated for the black-hole at rest in the  
center-of-mass frame of the colliding partons.  
The spectra are then transformed to the laboratory frame as needed.  
In all cases we have not (yet) taken the charge of the black-hole into account  
in calculating the emission spectrum, but have included phenomenological  
factors to account for it as explained below. 
 
\subsection{Electric and Color Charge Suppression} 
\label{subsec:qcsuppression} 
 
A charged and highly rotating black-hole will tend  
to shed its charge and angular momentum.  
Thus, emission of particles with charges  
of the same sign as that of the black-hole  
and angular momentum parallel to the black-hole's will be preferred.  
Emission of particles that increase the black-hole's charge or  
angular momentum should be suppressed.  
The precise calculation of these effects has not as yet been accomplished. 
Therefore, to account for these effects  
we allow optional phenomenological suppression factors  
for both charge and angular momentum.\\  
\\ 
The following charge-suppression factors can currently be used 
by setting parameter\\ 
{\tt charge$\_$suppression} (cf. section \ref{sec:input_output}) equal to 2. 
\begin{eqnarray} 
\label{eqn:highqandcsuppression} 
F^{Q}&=& \exp(\zeta_Q Q^{bh}Q^{em})\\ 
F^{3}_a&=& \exp(\zeta_3 c^{bh}_{a}c^{em}_{a}) \quad \quad {\rm a=r,b,g} . 
\end{eqnarray} 
$Q^{bh}$ is the electromagnetic charge of the black-hole,  
$Q^{em}$ is the charge of the emitted particle;  
$c^{bh}_{a}$, is the color value for the color $a$, with ${\rm a=r,b,g}$, of  
the black-hole,  
and 
$c^{em}_{a}$, is the color value for the color $a$, with ${\rm a=r,b,g}$, of the  
emitted particle.  
$\zeta_Q$ and $\zeta_3$ are phenomenological suppression parameters that are set  
as input parameters of the generator.\\ 
\\ 
We estimate  
$\zeta_{Q} = {\cal O}(\alpha_{em})$ and  
$\zeta_{3} = {\cal O}(\alpha_{s})$, 
where $\alpha_{em}$ and $\alpha_{s}$  
are the values of the electromagnetic and 
strong couplings at the Hawking temperature of the black-hole. 
Note that we currently neglect the possible restoration  
of the electroweak symmetry in the vicinity  
of the black-hole when its Hawking temperature 
is above the electroweak scale. 
Clearly, since $\alpha_{em}\simeq 10^{-2}$ we do not expect 
electromagnetic (or more correctly) electroweak 
charge suppression to be a significant effect. 
However, since $\alpha_s(1\,\rm{TeV})\simeq0.1$, 
color suppression may well play a role in the 
evolution of the black-hole. 
 
\subsection{Angular Momentum Suppression} 
\label{subsec:lsuppression} 

Since the TeV black-holes are quantum black-holes,  
the gray-body factors should really depend on both the initial 
and final black-hole parameters. The calculation of the gray-body spectra  
on a fixed background can cause some problems. 
In particular, in the current case,  the angular momentum of the  
emitted particle (as indeed the energy) may well be comparable  
to that of the black-hole itself.  There should be a 
suppression of particle  emission processes in which the black-hole  
final state is very different from the initial state.  
We therefore introduce a new phenomenological suppression factor, parameter {\tt L$\_$suppression},  
to reduce the probability of emission in which the angular momentum 
of the black-hole changes by a large amount.\\
\\ 
If parameter {\tt L$\_$suppression} is equal to 1  
(cf. section \ref{sec:input_output}), BlackMax does suppress the increase the angular momentum  
of the black-holes. 
There are three angular momentum suppression models BlackMax is able to simulate. 
\begin{enumerate}
\item $\Delta$Area suppression:\\ 
If the user sets {\tt L$\_$suppression} equal to 2 the $\Delta$Area suppression model is used where the suppression factor is defined as 
\be 
\label{eqn:LsuppressiondeltaA} 
F^{L}= \exp(\zeta_L (r_{h}^{bh}(t+\Delta t)^{2}/r_{h}^{bh}(t){^2}-1)), 
\ee 
here $\Delta t$ is the next time step in the simulation.
\item $J_{bh}$ suppression:
By setting {\tt L$\_$suppression} equal to 3 the $J_{bh}$ suppression model is used and the angular suppression factor is defined as
\be 
\label{eqn:LsuppressionJ} 
F^{L}= \exp(-\zeta_L |J^{bh}(t+\Delta t)|) . 
\ee 
\item $\Delta{J}_{bh}$ suppression:
For the $\Delta{J}_{bh}$ suppression model {\tt L$\_$suppression} has to be equal to 4 and 
\be 
\label{eqn:LsuppressiondeltaJ} 
F^{L}= \exp(-\zeta_L |{J}^{bh}(t+\Delta t)-J^{bh}(t)|) . 
\ee 
\end{enumerate}

We might expect $\zeta_L\sim1$, however there is no 
detailed theory to support this; as indeed there is no 
detailed theory to choose among these three phenomenological 
suppression scenarios.

 
\section{Final Burst of Black-Holes in BlackMax}
\label{sec:finalburst}

In the absence of a self-consistent theory of quantum gravity, 
the last stage of the evaporation cannot be described accurately.
Once the mass of black-hole becomes close to the fundamental 
scale $M_{*}$, the classical black-hole solution can certainly not be used 
anymore.
We adopt a scenario in which the final stage of evaporation 
is a burst of particles which conserves energy,
momentum\footnote{Black holes are not elementary particles and do not
conserve spin and angular momentum separatly. Hence spin is not
separately conserved in BlackMax, only the total angular momentum is.}  
and all of the gauge quantum numbers. 
For definiteness, we assume the remaining black-hole will decay 
into the lowest number of Standard-Model particles that conserve all quantum 
number, momentum and energy.\\
\\
Black-holes do not conserve global quantum numbers, like flavor,
baryon or lepton numbers~\cite{Stojkovic:2005zq}.   
If the user would like to conserve global quantum numbers s/he can set
this in the parameter-file (see section ~\ref{subsec:input}).

\section{Input and Output} 
\label{sec:input_output} 

\subsection{Input}
\label{subsec:input}

\begin{figure*}[htbp] 
\centering{ 
\includegraphics[height=8in]{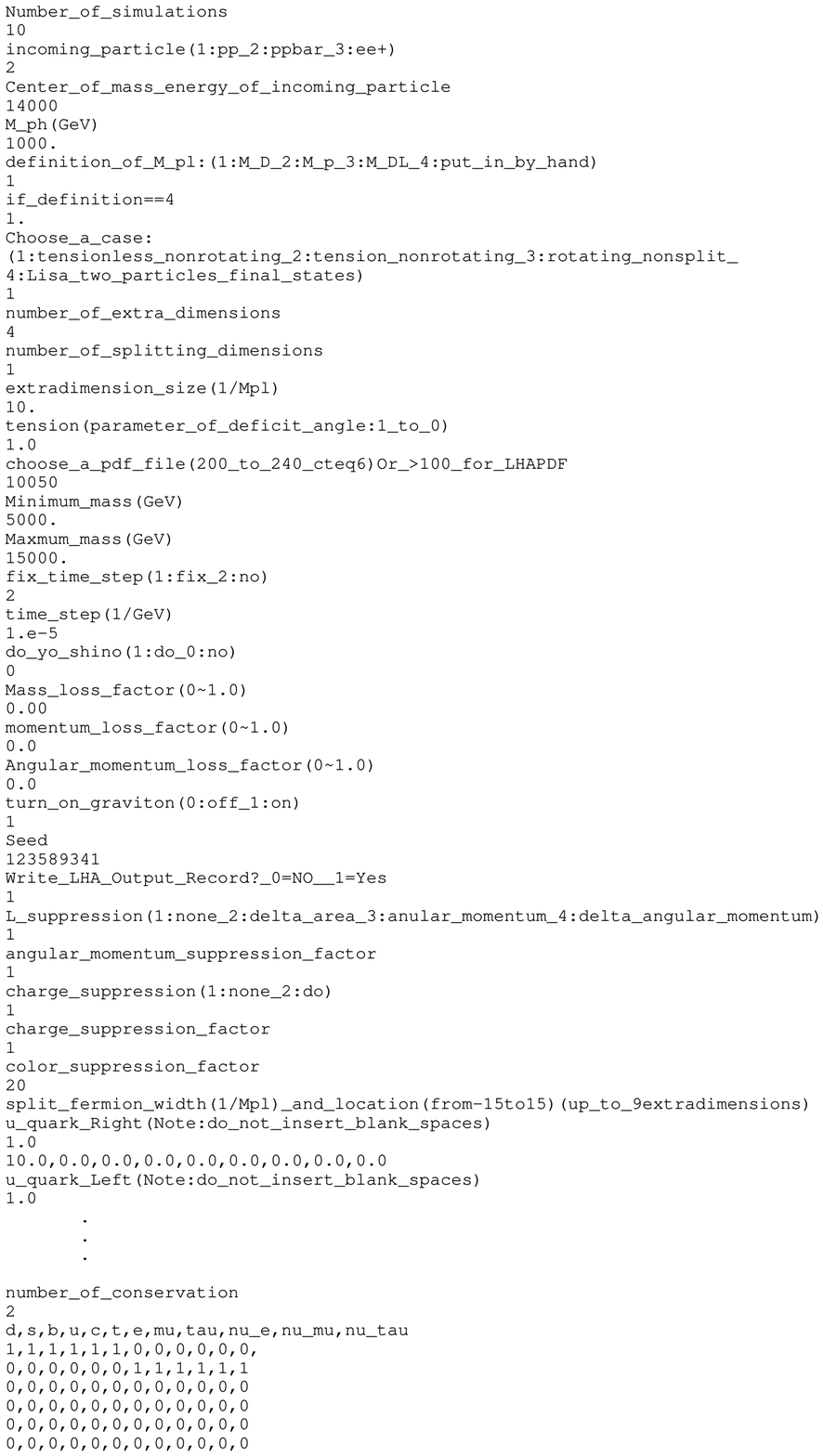} } 
\caption{parameter.txt is the input file containing the parameters that the user can set.} 
\label{fig:input} 
\end{figure*} 
 
The input parameters for the generator are read from the file parameter.txt, see  
Fig.\ref{fig:input}. In the following bulleted list an explanation is
given for each of the input parameters 
in parameter.txt.  

\begin{enumerate} 

\item {\tt Number$\tt{\_}$of$\tt{\_}$simulations}: 
sets the total number of  black-hole events to be simulated;

\item {\tt incoming$\_$particle(1:pp$\_$2:ppbar$\_$3:ee+)}: 
sets the type of incoming collision particles:
  \begin{itemize}
    \item 1 = proton-proton collisions,
    \item 2 = proton-antiproton collisions,
    \item 3 = electron-positron collisions.
  \end{itemize}

\item {\tt Center$\_$of$\_$mass$\_$energy$\_$of$\_$protons}: 
sets the center-of-mass energy of the colliding protons in GeV;

\item {\tt M$\_$ph}: 
sets the value of the fundamental quantum-gravity scale ($M_*$) in GeV;

\item {\tt definition$\_$of$\_$M$\_$pl:(1:M$\_$D$\_$2:M$\_$p$\_$3:M$\_$DL$\_$4:put$\_$in$\_$by$\_$hand)}: 
this sets the definition of the used fundamental quantum-gravity scale;
  \begin{itemize}
    \item 1 = Convention according to ~\cite{Abe:2001nqa} and first reference of \cite{Giudice:1998ck} (PDG definition).
    \item 2 = Convention according \cite{Peskin:2000ti}, which is useful in quoting experimental bounds (Giddings and Thomas).
    \item 3 = Convention given by Dimopoulos and Landsberg; 
             $M^{D-2}_{DL}=1/G_D$, with $D$ the total number of
             dimensions and $G_D$ the $D$-dimensional Newton gravity constant. 
    \item 4 = User defined convention. In this case BlackMax will read
the value of the next parameter {\tt if$\_$definition==4} to set the
scale. The value of {\tt if$\_$definition==4} is $k(d)$ in equation
\ref{eqn:horizon2}. Note that $k(d)$ is depending on the number of
total space dimensions and the user has to make sure to that the value of this
parameter is appropiately updated when the user simulates black
holes in scenarios with different number of extra dimensions. 
\end{itemize}	

\item {\tt Choose$\_$a$\_$case}: 
defines the extra dimension model to be simulated: 
\begin{itemize} 
\item 1 = non-rotating black-holes on a tensionless brane with  
possibility of fermion splitting,  
\item 2 = non-rotating black-holes on a brane with non-zero positive  
tension, 
\item 3 = rotating black-holes on a tensionless brane with d=5, 
\item 4 = two-particle final-state scenario; 
\end{itemize} 

\item {\tt number$\_$of$\_$extra$\_$dimensions}:
sets the number of extra dimensions; this must equal 2 for branes with non-zero positive tension ({\tt  
Choose$\_$a$\_$case}=2);
 
\item {\tt number$\_$of$\_$splitting$\_$dimensions}: \\
sets the number of extra split-fermion dimensions ({\tt Choose$\_$a$\_$case}=1); 

\item {\tt extradimension$\_$size}: 
sets the size of the mini-bulk\footnote{This is the distance between fermion branes where only gauge  
bosons and Higgs field can propagate in split-fermion brane scenario.} in units  
of $1/\rm{TeV}$ ({\tt Choose$\_$a$\_$case}=1); Please also refer to the discussion under item number 28. 

\item {\tt tension}:  
sets the deficit-angle parameter $B$~\cite{NKD,DKSS}; typical values
for branes with tension are from 1 to 0.9.  
({\tt Choose$\_$a$\_$case}=2);  

\item {\tt choose$\_$a$\_$pdf$\_$file(200$\_$to$\_$240$\_$cteq6)Or$\_>$100$\_$for$\_$LHAPDF}:  
The user can choose to run BlackMax with the bundled CTEQ6.1m PDFs~\cite{Stump:2003yu,cteqweb}
which come with the BlackMax release (see subdirectory cteq$\_$pdf)
or to run BlackMax with the PDFs of the LHAPDF package. 
In case the user is running BlackMax in standalone mode s/he has to
set this variable in parameter.txt between 200 and 240 to choose one
of the 41 CTEQ6.1M PDF sets. It is recommended to use 200 - the others
correspond to PDF uncertainties. 
If the user wants to use the PDFs of the LHAPDF package the
user must link the BlackMax executable to the LHAPDF library (see
section~\ref{subsec:lhapdf}). The input parameter to be given here
then is the chosen LHAPDF ID number. The definition of the LHAPDF
ID numbers can be found at~\cite{weblhapdf}. 

\item {\tt Minimum$\_$mass}: sets the minimum mass $M_{min}$  
in GeV of the initial black-holes; 

\item {\tt Maxmum$\_$mass}: sets the maximum mass $M_{max}$  
in GeV of the initial black-holes; 

\item {\tt fix$\_$time$\_$step}: 
\begin{itemize}
\item In case the user has choosen to simulate black-holes in a split
fermion scenario BlackMax will simulate the production and evaporation
of the black-holes in time steps. This variable influence the way the
size of the time steps are calculate in the split fermion scenario.
If set equal to 1, the code uses the   
parameter {\tt time$\_$step} to determine the time interval between
events; if set equal to 2 then code tries 
to optimize the time step, keeping the probability of emitting a particle 
in any given time step below 10\%.

\item In case the user has choosen to simulate black-holes in a {\bf
non}-split fermion scenario BlackMax will simulate the production and
evaporation of black-holes in fix time steps if its set to equal 1
using the input of {\tt time$\_$step} to set the size of the time
steps. If the user would like to save computation time s/he can turns
this off by setting this parameter to equal 2. In this case the code will
not calculate the location of the
black-hole for each time step, which speeds up the event generation.
\end{itemize}

\item {\tt time$\_$step}: 
defines the time interval $\Delta t$ in  
$\rm{GeV}^{-1}$ which the generator will use for the black-hole
evolution if ${\tt fix\_time\_step}=1$; 

\item {\tt do$\_$yo$\_$shino(1:do$\_$0:no)}:
If the user would like to include the Yoshino-Rychkov cross section enhancement 
factors and the energy loss before the event horizon fomation~\cite{Yoshino:2005hi} 
this parameter needs to be set equal 1 otherwise it should be
set to 0. In case it is set to 1 the {\tt Mass$\_$loss$\_$factor},
{\tt momentum$\_$loss$\_$factor} and {\tt
Angular$\_$momentum$\_$loss$\_$factor} will be ignored by BlackMax.

\item {\tt Mass$\_$loss$\_$factor}:  
sets the energy loss factor $0 \ge f_E \ge 1$ as defined in equation\,\ref{eqn:initlossfactors}; Recommended values
are between 10\% to 15\%. This depends also on the mimimum black-hole
mass the user wished to simulate. If the energy loss factor is too big
the probability to create a black-hole will be to low and BlackMax
would stop the generation. 

\item {\tt momentum$\_$loss$\_$factor}: defines the loss factor $0\leq  
f_p\leq 1$ for the momentum of initial black-holes as defined in  
equation\,\ref{eqn:initlossfactors}; Recommended values
are 10\% to 15\%.  

\item {\tt Angular$\_$momentum$\_$loss$\_$factor}: sets the loss factor $0  
\leq f_L \leq 1$ for the angular momentum of initial black-holes s defined in  
equation\,\ref{eqn:initlossfactors}; Recommended values
are 10\% to 15\%. 

\item {\tt turn$\_$on$\_$graviton(0:off$\_$1:on)}:
If this parameter is set to equal 1 BlackMax produces gravitons. Otherwise, 
BlackMax does not produce any gravitons. 

\item {\tt Seed}: 
This sets the seed for the random-number generator  
(9 digit positive integer); 

\item {\tt Write$\_$LHA$\_$Output$\_$Record(0$=$NO, 1$=$Yes, 2$=$more detailed output)}:
If the user sets this variable to 1 BlackMaxLHA.txt is written containing the full event info
and output.txt will contain input parameters and cross section
information only. If the user sets this variable 
equal to 2 BlackMaxLHA.txt is written containing full event information and output.txt will contain information
about the chosen input parameters, the cross section, the particles
which were emitted and the black-holes produced (see section~\ref{subsec:output}). 

\item {\tt L$\_$suppression}:  
This chooses the model for suppressing  
the accumulation of large black-hole angular momenta  
during the evolution phase of the black-holes  
(cf. discussion surrounding equations  
\ref{eqn:LsuppressiondeltaA}-\ref{eqn:LsuppressiondeltaJ}); 
\begin{itemize} 
\item 1 = no suppression;  
\item 2 = $\Delta$\,Area suppression;  
\item 3 = $J_{bh}$ suppression;  
\item 4 = $\Delta\,J$ suppression; 
\end{itemize} 

\item {\tt angular$\_$momentum$\_$suppression$\_$factor}: defines the  
phenomenological angular-momentum suppression factor, $\zeta_L$ (cf. discussion  
surrounding equation  
\ref{eqn:LsuppressiondeltaA}-\ref{eqn:LsuppressiondeltaJ}); Recommended value
is 0.2. 
 
\item {\tt charge$\_$suppression}: turns the suppression of accumulation of  
large black-hole electromagnetic and color charge during the black-hole  
evolution process on or off (cf. dicussion surrounding equation  
\ref{eqn:highqandcsuppression}) 
\begin{itemize} 
\item 1 = charge suppression turned off;  
\item 2 = charge suppression turned on; 
\end{itemize} 

\item {\tt charge$\_$suppression$\_$factor}:  
sets the electromagnetic charge suppression factor,  
$\zeta_Q$, in equation \ref{eqn:highqandcsuppression}; Recommended
values is 0.2. 

\item {\tt color$\_$suppression$\_$factor}:  
sets the color charge suppression factor,  
$\zeta_3$ in equation \ref{eqn:highqandcsuppression}; Recommended
values is 0.2. 

\item {\tt split$\_$fermion$\_$width(1/Mpl)$\_$and$\_$location(from-15to15)(up$\_$to$\_$9extradimensions)}  
In the first line after the name of the fermion, e.g. {\tt u$\_$quark$\_$Right}, the user can 
set the width of fermion wave functions (in $M_*^{-1}$ units). 
In the next line the user can set the centers of fermion wave functions (in $M_*^{-1}$ units) 
in split-brane models, represented as 9-dimensional vectors  
(for non-split models, set all entries to 0).

To satisfy the suppression of proton decay, the quark brane and
lepton brane must be separated by at least $10W$  (with $W$ the width
of fermion wavefunction in extra dimension) for one extra dimension. 
For higher extra dimensions, the separation can be smaller. It can be
less than $10W/n$ with $n$ the number of extra dimension. 

To suppress n-nbar oscillation, the quarks have to be
separated. The recommended value needs to be at least $3W/n$. to
$5W/n$, with $n$ the number of extra dimensions. 


There is no way to satisfy all the constraints at the same time for
the split-fermion case. But there are several papers in which the
possible position for each fermion from the CP violation constraint
are calculated. \cite{Mirabelli:1999ks,Branco:2000rb} which we would
like to refer the user to.

\item {\tt number$\_$of$\_$conservation}:
During the balding and evaporation
phase, the black hole will emit a certain number of quarks and leptons
and acquire global quantum numbers~\cite{Stojkovic:2005zq}.\footnote{It will also acquire angular
momentum and electric and color charges, but this is not relevant for
this discussion.}  These quantum numbers are recorded by BlackMax as it
reaches the final burst step, at which point it will generate further
particle emissions so that the whole process, from beginning to end,
will obey certain user-specified global fermion number conservation rules.

The {\tt number$\_$of$\_$conservation} parameter specifies the number of
conservation rules which are to follow the subsequent
{\tt d,s,b,u,c,t,e,mu,tau,nu$\_$e,nu$\_$mu,nu$\_$tau} line in the
parameter file. 
The individual conservation rules are specified as an ordered set of
integral
coefficients $\{a_f\}$ (conservation matrix) such that the quantity $\sum_f a_fN_f$ is a constant,
where $N_f$ is the number of particular fermion flavor.

For example, baryon number $\rm{B}$ can be conserved in BlackMax by making sure
that in the final burst stage, the sum of the number of emitted $u$
and $d$-type quarks compensate for the number $u$ and $d$-type quarks
emitted in previous stages.  The conservation rule is thus
\begin{equation}
  N_u + N_d + N_c + N_s + N_t + N_b = 3\rm{B}
\end{equation}
(recalling that quarks have a baryon number of 1/3).  If this is the only
additional rule, then the following lines would be entered into the parameter
file:
\begin{verbatim}
  number_of_conservation
  1
  d,s,b,u,c,t,e,mu,tau,nu_e,nu_mu,nu_tau
  1,1,1,1,1,1,0,0,0,0,0,0
\end{verbatim}
The zeroes for the leptons indicate that they play no role in baryon number
conservation.
If, on the other hand, the user wishes to conserve lepton number $\rm{L}$, the
quark and lepton coefficients are swapped:
\begin{verbatim}
  number_of_conservation
  1
  d,s,b,u,c,t,e,mu,tau,nu_e,nu_mu,nu_tau
  0,0,0,0,0,0,1,1,1,1,1,1
\end{verbatim}
The two rules can be combined in the parameter file to conserve $\rm{B}$ and $\rm{L}$
separately:
\begin{verbatim}
  number_of_conservation
  2
  d,s,b,u,c,t,e,mu,tau,nu_e,nu_mu,nu_tau
  1,1,1,1,1,1,0,0,0,0,0,0
  0,0,0,0,0,0,1,1,1,1,1,1
\end{verbatim}

Conservation of $\rm{B-L}$ takes the following form:
\begin{equation}
  \frac{1}{3}(N_u + N_d + N_c + N_s + N_t + N_b) -
  (N_e + N_{\nu_e} + N_\mu + N_{\nu_\mu} + N_\tau + N_{\nu_\tau}) = \rm{B} - \rm{L}
\end{equation}
from which parameters follow:
\begin{verbatim}
  number_of_conservation
  1
  d,s,b,u,c,t,e,mu,tau,nu_e,nu_mu,nu_tau
  1,1,1,1,1,1,-3,-3,-3,-3,-3,-3
\end{verbatim}

Finally, if all fermion flavors are to be conserved individually, twelve
equations need to be specified:
\begin{verbatim}
  number_of_conservation
  12
  d,s,b,u,c,t,e,mu,tau,nu_e,nu_mu,nu_tau
  1,0,0,0,0,0,0,0,0,0,0,0
  0,1,0,0,0,0,0,0,0,0,0,0
  0,0,1,0,0,0,0,0,0,0,0,0
  0,0,0,1,0,0,0,0,0,0,0,0
  0,0,0,0,1,0,0,0,0,0,0,0
  0,0,0,0,0,1,0,0,0,0,0,0
  0,0,0,0,0,0,1,0,0,0,0,0
  0,0,0,0,0,0,0,1,0,0,0,0
  0,0,0,0,0,0,0,0,1,0,0,0
  0,0,0,0,0,0,0,0,0,1,0,0
  0,0,0,0,0,0,0,0,0,0,1,0
  0,0,0,0,0,0,0,0,0,0,0,1
\end{verbatim}

It should be noted that not all the possibilities the user can enter as
rules are realized in BlackMax.  If the user specifies a rule which
BlackMax cannot implement, it will print an error message and halt.

Figure~\ref{fig:input} shows an example where {\tt
number$\_$of$\_$conservation} is set to 2 and there are 6 lines in
the conservation matrix. Only the first two lines of the conservation
matrix in Figure~\ref{fig:input} will be read and the rest will be ignored.

\end{enumerate}

\subsection{Output}
\label{subsec:output}

The BlackMax code produces 3 types of output: basic information
printed to screen, the BlackMaxLHArecord.txt file, and the output.txt
file shown in Figs. 2, 3, and 4. This is controlled by the parameter
{\tt Write$\_$LHA$\_$Output$\_$Record} which can be set to 0, 1, or 2 for
increasingly detailed output.  \\

The screen dump provides basic version and input settings, cross
section and timing information and can be redirected to a file. This
information is always given. For parameter value 0 output.txt is
written and includes a complete list of all input parameters and the
calculated cross section only. For parameter value 1 (default) the
BlackMaxLHArecord.txt file is additionally created containing all
input parameters, followed by an event-by-event record in LHA
format~\cite{Boos:2001cv}. This file may be used as input to 
any other LHA compliant MC e.g. for further hadronisation by Pythia or
HERWIG. Finally for parameter value 2 the output.txt file then also
contains much more detailed event-by-event information for each
emission step in the black-hole decay. The information is tagged by an
ID word (Begin,Parent, Pbh, trace, Pem, Pemc or Elast). Note that the
energy in the output.txt file 
is the bulk energy of the particles where as the energy in
BlackmaxLHArecord.txt is the observed energy as defined in equation 57 in ~\cite{Dai:2007ki}.

\begin{figure*}[htbp] 
\includegraphics[width=6in]{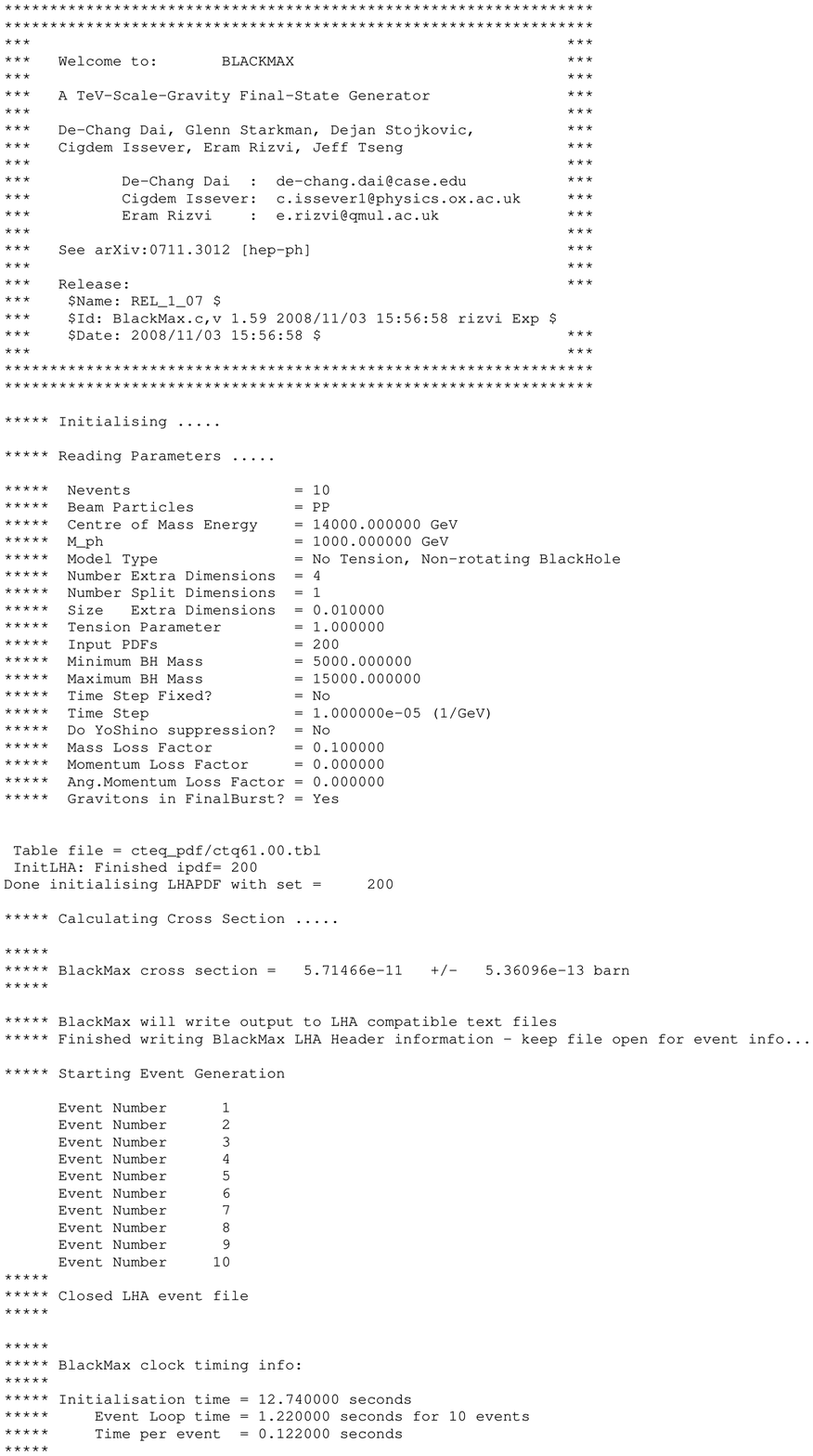}
\caption{This is the basic information which is printed to screen
and which can be redirected to a file. It contains information about
version numbers, dates, input parameters and run time.}  
\label{fig:out}  
\end{figure*} 

\begin{figure*}[htbp] 
\includegraphics[trim = 0mm 50mm 0mm 0mm, clip,width=1.0\textwidth]{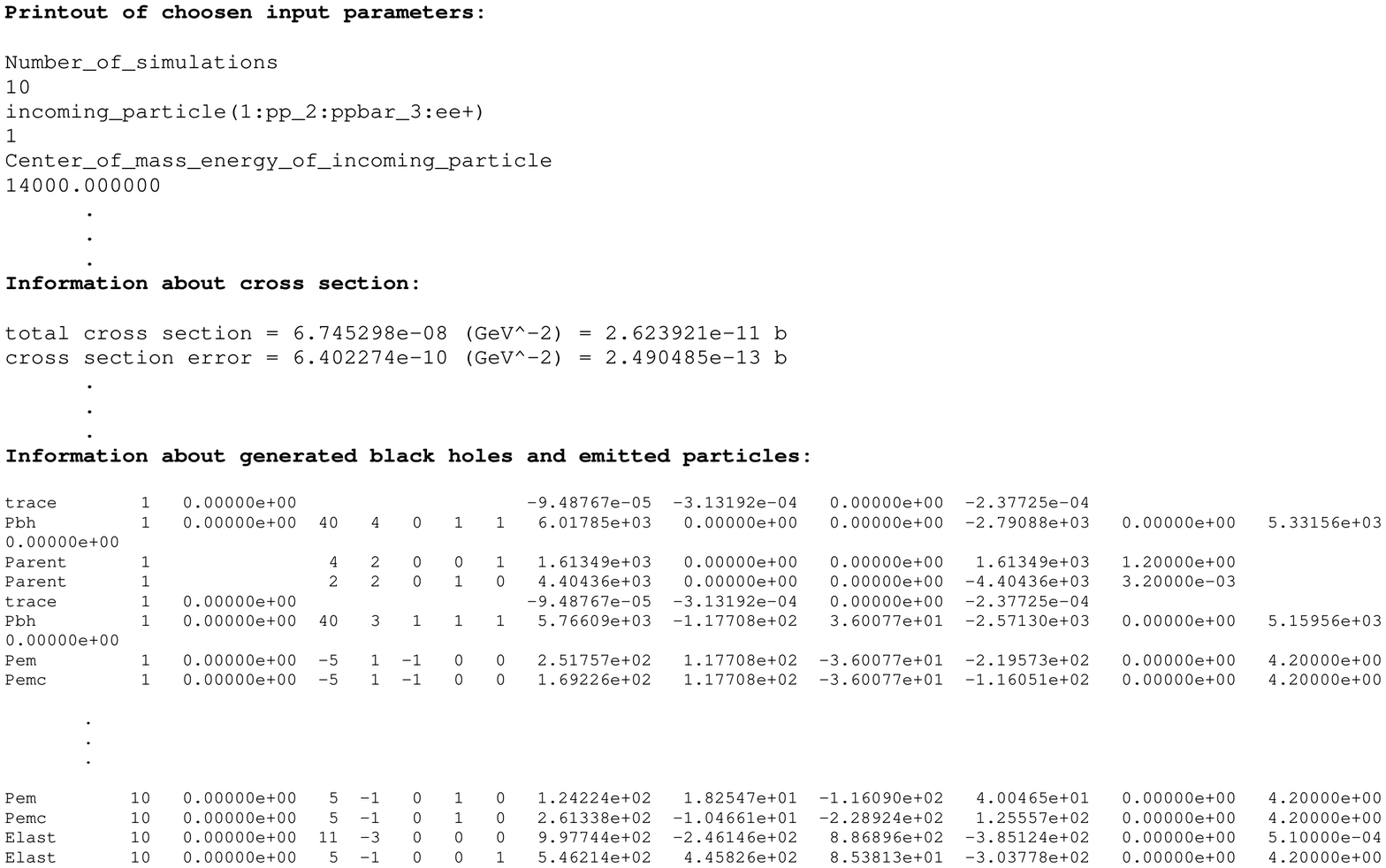}
\caption{output.txt: There are three parts to this file.  
The first part is a copy of parameter.txt followed by the information
on the black-hole production cross section as inferred from the events
in this generator run and is produced for {\tt
Write$\_$LHA$\_$Output$\_$Record} equal to 0 and 1. 
The third part includes  information about the black-hole and the
emitted particles and is produced for {\tt
Write$\_$LHA$\_$Output$\_$Record} equal to 2.  
The first column identifies what type of information each row is supplying:
lines starting with ``Begin'' contain information about the emitted
particles before the formation of the black-hole; lines beginning with
``parent'' have information about the two incoming partons; lines
beginning ``Pbh'' contain information on the
energy and momenta of the produced black-holes;  
lines starting with ``trace'' describe the location of the black-holes; 
rows beginning with ``Pem'' contain information about the emitted particles in the lab frame; 
lines headed by ``Pemc'' have the information about the emitted particles in the center-of-mass frame;  
rows starting with ``Elast'' describe the final burst.} 
\label{fig:screenoutput}  
\end{figure*} 

\begin{figure*}[htbp] 
\includegraphics[trim = 0mm 50mm 0mm 0mm, clip,width=6.5in]{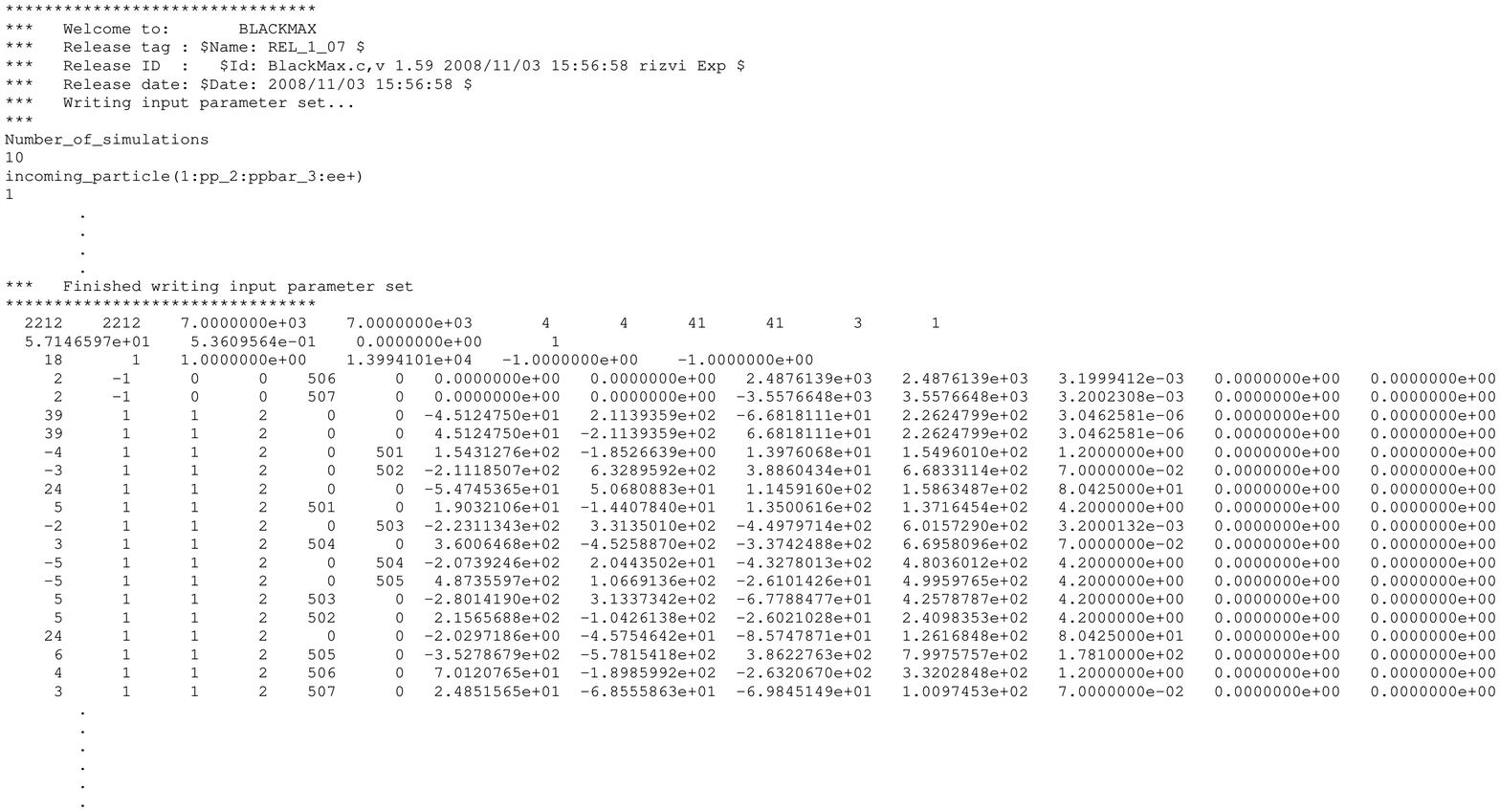}
\caption{BlackMaxLHArecord.txt: This file has the output of BlackMax
in LHA format.} 
\label{fig:LHAout}  
\end{figure*} 
 
\begin{figure*}[htbp] 
\centering{ 
\includegraphics[trim = 25mm 170mm 0mm 10mm, clip,width=7.5in]{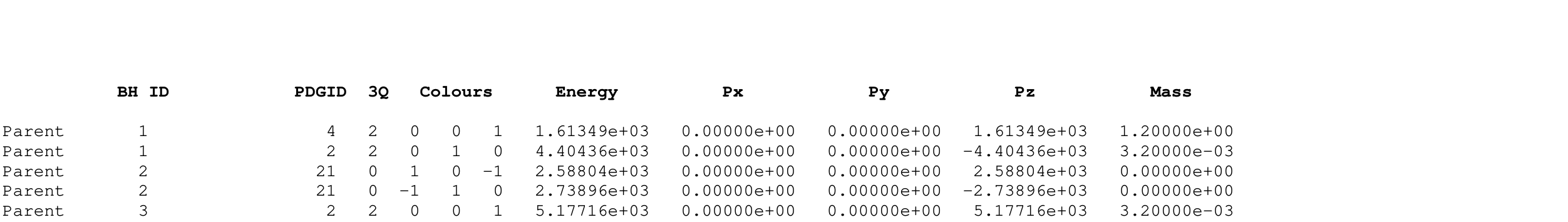} } 
\caption{Lines in the output file headed by the ID = {\tt Parent} contain  
information about  
the initial partons which formed the black-hole.} 
\label{fig:output1} 
\end{figure*} 

\begin{figure*}[htbp] 
\centering{ 
\includegraphics[trim = 10mm 170mm 0mm 10mm, clip,width=6.5in]{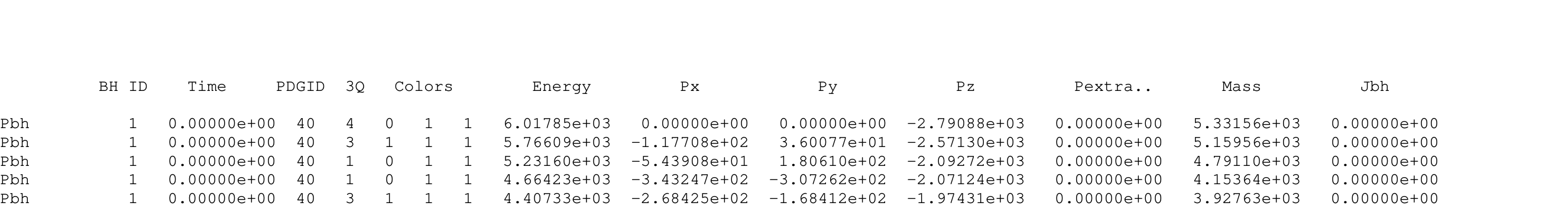} } 
\caption{Lines in the output file headed by the ID = {\tt Pbh} contain the  
energies and momenta  
of the black-holes for each emission step. In case of rotating black-holes,  
the last column in the line is the angular momentum.} 
\label{fig:output2} 
\end{figure*} 

\begin{figure*}[hb] 
\centering{ 
\includegraphics[trim = 25mm 170mm 0mm 10mm, clip,width=7.5in]{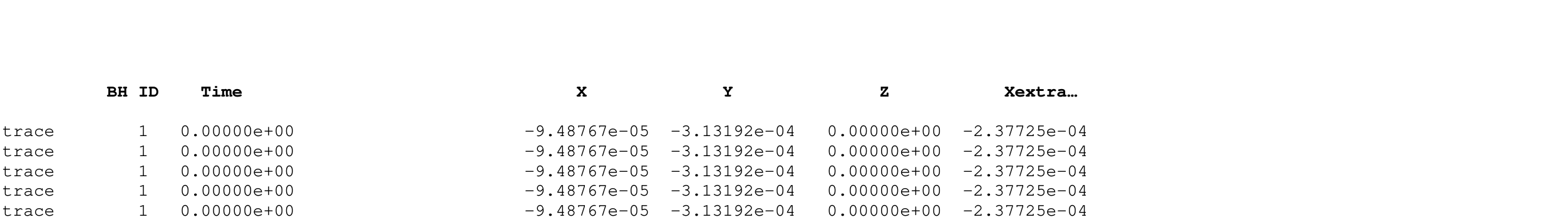} } 
\caption{Lines in the output file headed by the ID = {\tt trace} contain the  
location  
of the black-hole fo reach emission step.} 
\label{fig:output3} 
\end{figure*} 

\begin{figure*}[htbp] 
\centering{ 
\includegraphics[trim = 25mm 170mm 0mm 10mm, clip,width=7.5in]{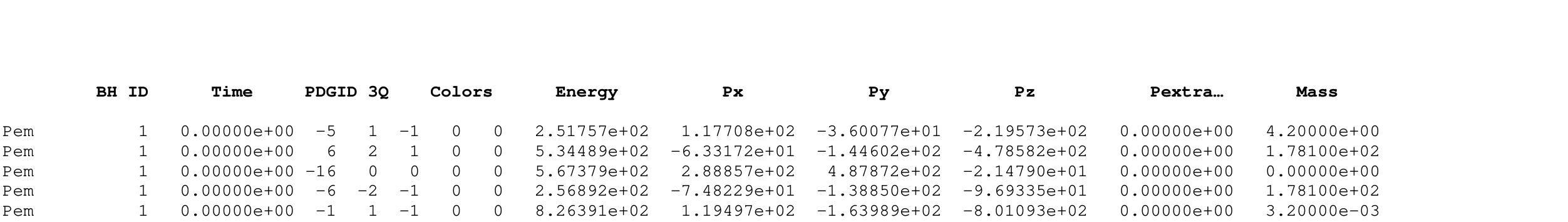} } 
\caption{Lines in the output file headed by the ID = {\tt Pem} contain the  
types of the emitted particles, their energies and momenta in the lab frame  
and the times of their emission.} 
\label{fig:output45} 
\end{figure*} 

\begin{figure*}[htbp] 
\centering{ 
\includegraphics[trim = 25mm 170mm 0mm 10mm, clip,width=7.5in]{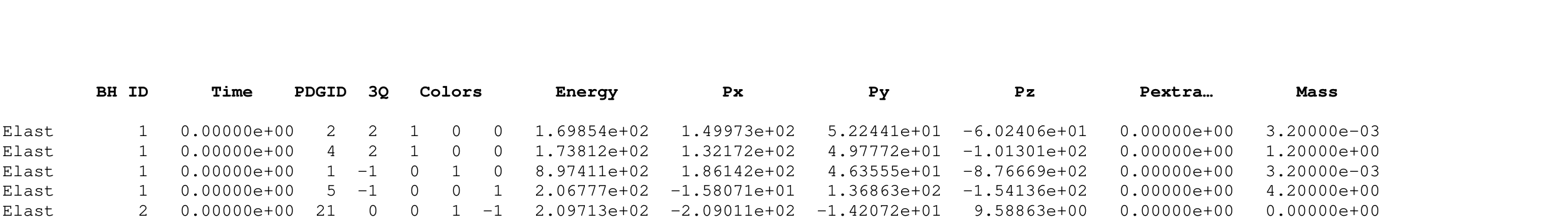} } 
\caption{Lines in the output file headed by the ID = {\tt Elast} contain the  
types, energies and momenta of particles of the final burst.} 
\label{fig:output6} 
\end{figure*} 
 
\begin{itemize} 
\item {\bf Parent}: identifies the partons whose collision resulted in the  
formation of the initial black-hole (see Fig. \ref{fig:output1}). 
\begin{itemize} 
\item column 1: identifies the black-hole; 
\item column 2: PDGID code of the parton;  
\item column 3: electric charge of parent parton in $3Q$;
\item column 4-6: color-charge vector  
components of the parent parton; 
\item column 7: energy of the parton in GeV; 
\item columns 8-10: brane momenta of the parton in GeV;
\item colunm:11: mass of the parton in GeV. 
\end{itemize} 
\item {\bf Pbh}: contains the evolution  
of the charge, color, momentum and energy of the black-holes,  
and, for rotating black-holes, their angular momentum  
(cf. Fig. \ref{fig:output2}). 
\begin{itemize} 
\item column 1: identifies the black-hole; 
\item column 2: time at which the black-hole  
emitted a particle; 
\item column 3: PDGID code of a black-hole; 
\item column 4: three times the  
electromagnetic charge of the black-hole; 
\item columns 5 to 7: color-charge vector  
components of the black-hole; 
\item columns 8: energy of the black-hole  
in the laboratory frame; 
\item columns 9 to 11: brane components  
of the black-hole momentum in the laboratory frame; 
\item columns 12 to (8+d): bulk components  
of the black-hole momentum; 
\item column (9+d): mass of the black-hole;
\item column (10+d): angular momentum  
of the black-hole,  in the case of rotating black-holes;  
empty otherwise. 
\end{itemize} 
\item {\bf trace}: contains the evolution history  
of the black-holes' positions (cf. Fig. \ref{fig:output3}): 
\begin{itemize} 
\item column 1: identifies the black-hole;  
\item column 2: the times at which the black-hole  
emitted a particle; 
\item columns 3 - 5 are the brane components  
of the black-hole position vector 
when the black-hole emitted a particle; 
\item columns 6 to (2+d): the bulk components  
of the black-hole position vector, 
when the black-hole emitted a particle. 
\end{itemize} 
\item {\bf Pem}: contains a list of particles the black-holes has emitted during its evaporation phase (cf. Fig. \ref{fig:output45}): 
\begin{itemize} 
\item column 1: identifies the black-hole; 
\item column 2: the times at which the  
black-hole emitted a particle; 
\item column 3: PDGID code of the emitted particle; 
\item column 4: three times  
the charge of the emitted particle; 
\item columns 5 to 7: color-vector components  
of the emitted particle; 
\item columns 8: energy of the emitted particle  
in the laboratory frame in GeV; 
\item columns 9 to 11: brane components  
of the momentum of the emitted particle, 
in the laboratory frame in GeV; 
\item columns 12 to (8+d): bulk components  
of the momentum of the emitted particle in GeV; 
\item column 9+d: mass of the emitted particle in GeV.
\end{itemize} 
\item {\bf Pemc}: contains the same information as Pem, 
but in the center-of-mass frame of the collision. 
\item {\bf Begin}: contains the same information as Pem, 
but for the particles which are emitted before the black-hole is
formed. This happens if the user chooses to use the Yoshion and Rykhov
factors or sets the mass-loss-factor to a non-zero value. 
\item {\bf Elast}: contains the same information as Pem 
for the particles emitted in the final decay burst of the black-hole but 
column 12 is the mass of the particle. The is no information on the bulk 
momentum since these particles have no bulk momentum.  
\end{itemize}




\appendix
\section{Comparison between BlackMax and Charybdis}
\label{app:comparison}

In this section we compare the cross sections of BlackMax and
Charybdis~\cite{Harris:2003db} which is a very commonly used black
hole generator\footnote{For the comparison in this section MSRT98LO
was used as the input PDF.}. 
There are several differences between BlackMax and Charybdis which needs to be considered.\\
\\
For a direct comparison one has to make sure that both generators use
the same convention for the definition of the Planck Mass. In Charybdis
the user can choose between the different conventions by setting the parameter
{\tt MSSDEF}. In BlackMax the user can switch between the different
conventions with the help of the parameter {\tt
definition$\_$of$\_$M$\_$pl}. 

\begin{itemize}
\item {\bf PDG definiton, $M_D$:} \\ {\tt definition$\_$of$\_$M$\_$pl} = 1 corresponds
to {\tt MSSDEF} = 3; 
\item {\bf Giddings and Thomas, $M_{p}$:} \\ {\tt definition$\_$of$\_$M$\_$pl} = 2 corresponds
to {\tt MSSDEF} = 1; 
\item {\bf Dimopoulos and Landsberg, $M_{DL}$:} \\  {\tt definition$\_$of$\_$M$\_$pl} = 3 corresponds
to {\tt MSSDEF} = 2.
\end{itemize}

One important difference between the two generators is the definition
of cross sections. Charybdis uses the cross section definition for a
non-rotating black-hole 
\begin{equation}
\sigma_{ch}= \pi r_s^2.
\end{equation}
Here, $r_s$ is the Schwarzchild radius of the black-hole. 

But BlackMax uses 
\begin{eqnarray}
\sigma_{bm}&=&b_d^2\pi r_s^2\\
b_d&=&\frac{2}{(1+(\frac{d-1}{2})^2)^\frac{1}{d-2}}
\end{eqnarray}
which is the definiton of the cross section for rotating
black-holes (see also equation (\ref{bmax2})),  
where d is number of space dimensions. The variable $b_d$ is in general larger than one. Therefore one expects that the 
cross section of BlackMax is larger than the cross section of Charybdis. $\sigma_{bm}/\sigma_{ch}$ is 
expected to be equal to $b_d^2$. In the tables~\ref{MD} to~\ref{mdl}  the
value of $\sigma_{bm}/\sigma_{ch}$ is in general smaller than $b_d^2$,
and in some cases it is even smaller than 1. 
This difference comes from the fact that BlackMax assumes 
that the width of the extradimension is $M_{pl}^{-1}$, and
Charybdis assumes that the width of extradimension is 0. 
Because of the finite width of extradimensions, the cross section of
BlackMax is further reduced compared to Charybdis.\\
\\
To make sure that the two generators indeed output the same cross
section in the same situation, we manually put the size of extradimension to 
zero in BlackMax for the purpose of this comparison. 
The results are shown in the sixth column of the
tables~\ref{MD} to ~\ref{mdl} and they agree with the expected value well and
there is a small difference between $b_d^2$ and
$\sigma_{bm}/\sigma_{ch}$ of 3$\%$.

\begin{table}[h]
\caption{$M_D$=1000 GeV, $M_{bh}>5000$ GeV, and D is the total number of dimensions (space + time).}
\label{MD}
\begin{tabular}{|c|c|c|c|c|c||c|}
\hline			
D & $\sigma_{ch}$ [pb] & $\sigma_{bm}$ [pb] & $\sigma_{bm}$ with $L_{extra}=0$ [pb] & $\sigma_{bm}/\sigma_{ch}$& $\sigma_{bm}/\sigma_{ch}$ with $L_{extra}=0$& $b_d^2$\\ 
\hline
6& 75.20 $\pm$  0.6968 &  90.69 $\pm$ 0.8407&99.70$\pm$  0.9128 & 1.21 & 1.32& 1.36\\
\hline
7&  122.0$\pm$ 1.126  &  161.9$\pm $   1.502& 177.0$\pm$1.638 &  1.32& 1.45 & 1.48\\
\hline
8& 172.6 $\pm $1.590 & 247.6 $\pm$  2.304& 266.2$\pm$ 2.449& 1.43&1.54 & 1.59\\
\hline
9& 225.7$\pm$ 2.076  &  352.7 $\pm $ 3.149&369.0$\pm$3.285 & 1.56&1.63 & 1.69\\
\hline
10& 280.7$\pm$  2.579 &  455.2$\pm$ 4.182&484.8$\pm$ 4.419& 1.62 & 1.72& 1.78\\
\hline  
\end{tabular}
\end{table}

\begin{table}[h]
\caption{$M_p$=1000 GeV, $M_{bh}>5000$ GeV, and D is total number of dimensions (space + time).}
\label{Mp}
\begin{tabular}{|c|c|c|c|c|c||c|}
\hline			
D & $\sigma_{ch}$ [pb] & $\sigma_{bm}$ [pb] & $\sigma_{bm}$ with $L_{extra}=0$ [pb]& $\sigma_{bm}/\sigma_{ch}$& $\sigma_{bm}/\sigma_{ch}$ with $L_{extra}=0$& $b_d^2$\\
\hline
6&  119.4$\pm$1.106   & 149.1  $\pm$ 1.375&158.3$\pm$ 1.449  & 1.25 &1.33 & 1.36\\
\hline
7&  172.6$\pm$1.593  &  234.5$\pm $2.158  & 250.4$\pm$ 2.316& 1.36 & 1.45 & 1.48\\
\hline
8&  227.8$\pm $2.098 &  334.2$\pm$3.090 & 351.2$\pm$3.231 &1.47 & 1.50& 1.59\\
\hline
9& 284.4$\pm$2.626  &  450.2$\pm $ 4.010&464.8$\pm$4.139&1.58 &1.63 & 1.69\\
\hline 
10& 342.1$\pm$3.144   &561.5  $\pm$5.186 &591.0$\pm$ 5.387&1.64 &1.73 & 1.78\\
\hline
\end{tabular}
\end{table}

\begin{table}[h]
\caption{$M_{DL}$=1000 GeV, $M_{bh}>5000$ GeV, and D is the total number of dimensions (space + time).}
\label{mdl}
\begin{tabular}{|c|c|c|c|c|c||c|}
\hline			
D & $\sigma_{ch}$ [pb] & $\sigma_{bm}$ [pb] & $\sigma_{bm}$ with $L_{extra}=0$ [pb] & $\sigma_{bm}/\sigma_{ch}$& $\sigma_{bm}/\sigma_{ch}$ with $L_{extra}=0$& $b_d^2$\\
\hline
6&  55.65$\pm$0.5156   &  64.82 $\pm$ 0.6047&73.78$\pm$ 0.6755  & 1.16 & 1.33& 1.36\\
\hline
7&  38.85$\pm$0.3586  &  41.78$\pm $ 0.3894 & 56.35$\pm$0.5212 & 1.08 & 1.45 & 1.48\\
\hline
8&  33.12$\pm $0.3050 &  32.86$\pm$0.3083 & 51.07$\pm$0.4698 & 0.992&1.54 & 1.59\\
\hline
9& 30.90$\pm$0.2842  &  27.81$\pm $0.2590 &50.51$\pm$ 0.4498& 0.900&1.63 & 1.69\\
\hline
10& 30.20$\pm$0.2775   &  24.70$\pm$0.2353 &52.16$\pm$0.4755 &0.818 &1.73 & 1.78\\
\hline  
\end{tabular}
\end{table}

\bibliographystyle{utphys}

\bibliography{refs}{}

\end{document}